# Toward a Broadband Astro-comb: Effects of Nonlinear Spectral Broadening in Optical Fibers

**Guoqing Chang[1*], Chih-Hao Li[2], David F. Phillips[2], Ronald L. Walsworth[2,3], and Franz X. Kärtner[1]**

[1]*Department of Electrical Engineering and Computer Science and Research Laboratory of Electronics, Massachusetts Institute of Technology, 77 Mass. Ave Cambridge MA 02139*
[2]*Harvard-Smithsonian Center for Astrophysics, Harvard University, 60 Garden St. Cambridge MA 02138*
[3]*Department of Physics, Harvard University, Cambridge MA 02138*

[*]*guoqing@mit.edu*

**Abstract:** We propose and analyze a new approach to generate a broadband astro-comb by spectral broadening of a narrowband astro-comb inside a highly nonlinear optical fiber. Numerical modeling shows that cascaded four-wave-mixing dramatically degrades the input comb's side-mode suppression and causes side-mode amplitude asymmetry. These two detrimental effects can systematically shift the center-of-gravity of astro-comb spectral lines as measured by an astrophysical spectrograph with resolution ≈100,000; and thus lead to wavelength calibration inaccuracy and instability. Our simulations indicate that this performance penalty, as a result of nonlinear spectral broadening, can be compensated by using a filtering cavity configured for double-pass. As an explicit example, we present a design based on an Yb-fiber source comb (with 1 GHz repetition rate) that is filtered by double-passing through a low finesse cavity (finesse = 208), and subsequent spectrally broadened in a 2-cm, SF6-glass photonic crystal fiber. Spanning more than 300 nm with 16 GHz line spacing, the resulting astro-comb is predicted to provide 1 cm/s (~10 kHz) radial velocity calibration accuracy for an astrophysical spectrograph. Such extreme performance will be necessary for the search for and characterization of Earth-like extra-solar planets, and in direct measurements of the change of the rate of cosmological expansion.

**OCIS codes**: (060.7140) Ultrafast processes in fibers; (120.6200) Spectrometers and spectroscopic instrumentation; (060.4370) Nonlinear optics, fibers.

---

## References and links

## 1. Introduction

Astro-combs, i.e., laser frequency combs optimized for wavelength calibration of astronomical spectrographs, are an enabling tool for precision radial velocity observations, including the search for and characterization of small, rocky (Earth-like) extra-solar planets (exoplanets), direct observation of cosmological deceleration, and the study of temporal variation of fundamental constants over cosmological time scales [1-4]. An astro-comb originates from a mode-locked femtosecond laser (known as the "source comb"), which produces a set of bright, evenly-spaced comb lines in the frequency domain; the source comb's line spacing is equal to the laser's repetition rate, typically ≤ 1 GHz. Such a small line spacing prevents current high resolution astrophysical spectrographs ($R=\lambda/\Delta\lambda$ = 10,000 - 100,000) from resolving the source comb's spectral features; i.e., the recorded spectrum would be a "white" continuum useless for calibration.

Direct increase of the line spacing requires increasing the pulse repetition rate of the mode-locked femtosecond laser (to >15 GHz), which in turn demands short cavities (cavity round-trip length < 2 cm), rendering mode-locking for femtosecond pulse generation extremely difficult. This dilemma has been solved by integrating the source comb with a stabilized Fabry-Perot (FP) cavity with a free-spectral-range (FSR) equal to $M$ times the source comb's repetition rate, where $M$ is an integer. Such a filtering cavity passes every $M^{th}$ source comb line and blocks intermediate lines ("side-modes" hereafter). Referenced to an atomic clock or to GPS (Global Positioning System), the resulting astro-comb has long term stability and reproducibility superior to any traditional wavelength calibrators, such as thorium-argon lamps and iodine absorption cells.

An ideal wavelength calibrator will provide spectral coverage across an astronomical spectrograph's full range, typically >300 nm. To date, however, the maximum astro-comb spectral coverage is only ~100 nm [2], primarily limited by the FP filtering cavity's dispersion both from the cavity mirrors and air. Cavity dispersion leads to a variation of the FP cavity FSR over different wavelength ranges. Consequently, frequency mismatch between the cavity transmission resonances and the source comb lines causes a variation in transmitted line intensity, which shifts the center-of-gravity of the astro-comb lines, and dramatically narrows the useful transmitted spectral width. A naive solution to the limited bandwidth of current astro-combs is to lower the FP cavity's finesse. However, this usually leads to larger shifts of the center-of-gravity of astro-comb lines. For example, for a 20-GHz astro-comb generated from a 1 GHz source comb and a FP cavity with moderate finesse of 200 (corresponding to 26 dB suppression of the least suppressed side modes), the dispersion from air limits the bandwidth of the astro-comb to <200 nm. Not only is the design of a FP cavity with low dispersion over a broad bandwidth particularly difficult, it is also challenging to design a FP cavity with high throughput and good mode-matching over a broad spectrum. In this paper, we propose and analyze a new approach to realize a broadband (>300 nm) astro-comb with relaxed requirements on the bandwidths of both the source comb and the FP filtering cavity.

## 2. Broadband astro-comb design

As illustrated in Fig. 1(a), the proposed broadband astro-comb consists of three main components: a source comb constructed from a mode-locked femtosecond laser, a narrowband FP cavity operating at double-pass configuration for repetition-rate multiplication, and a highly nonlinear optical fiber for spectral broadening.

The key enabling technique in the design is efficiently suppressing unwanted side-modes, which is achieved by double-passing the source comb through a FP filtering cavity. In contrast to the conventional single-pass configuration, double-pass filtering, albeit at the cost of a slight (few percent) decrease in transmission bandwidth, provides ultrahigh side-mode suppression with a low finesse FP cavity [5, 6]. For example, constructing a double-pass filtering cavity from two identical mirrors with 98.5% reflectivity (corresponding to a relatively low finesse of 208) and FSR of 16 GHz (designed to pass every $16^{th}$ line) provides suppression of the first and second side-mode of 56 dB and 68 dB, respectively [Fig. 1(b)]. By comparison, using single-pass filtering to achieve the same suppression of the first side-mode requires a FP cavity finesse >5,000 (equivalent to cavity mirrors with 99.94% reflectivity), which imposes extreme difficulty on practical implementation. Figure 2 compares calculated power transmission and phase for three different FP cavities: a low-finesse cavity at single-pass, low-finesse cavity at double-pass, and high-finesse cavity at single-pass. While the latter two schemes possess the same suppression for the first side-mode, the single-pass, high-finesse cavity suppresses higher-order side-modes less than is achieved by its double-pass counterpart. As the following sections will detail, stronger suppression of higher-order side-modes before nonlinear spectral broadening is crucial for a broadband astro-comb to achieve 1 cm/s (~10 kHz) calibration accuracy on astrophysical spectrographs.

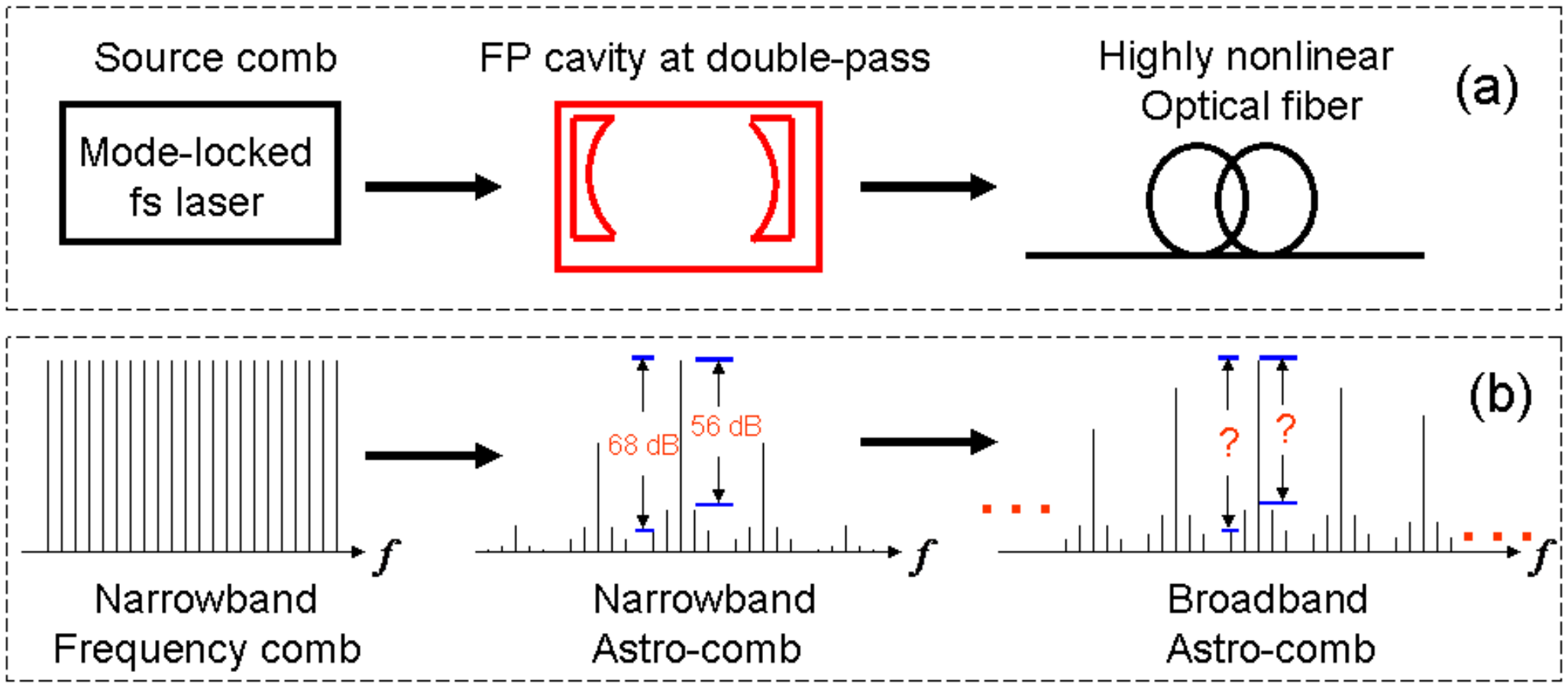

Fig. 1. (a) Proposed approach for a broadband astro-comb including 3 main components; and (b) corresponding frequency comb spectrum for each stage. 56 dB and 68 dB in (b) indicate suppression of the first and second side-mode as the source comb double passes a Fabry-Perot (FP) cavity constructed using two identical mirrors of 98.5% reflectivity (see main text for more details).

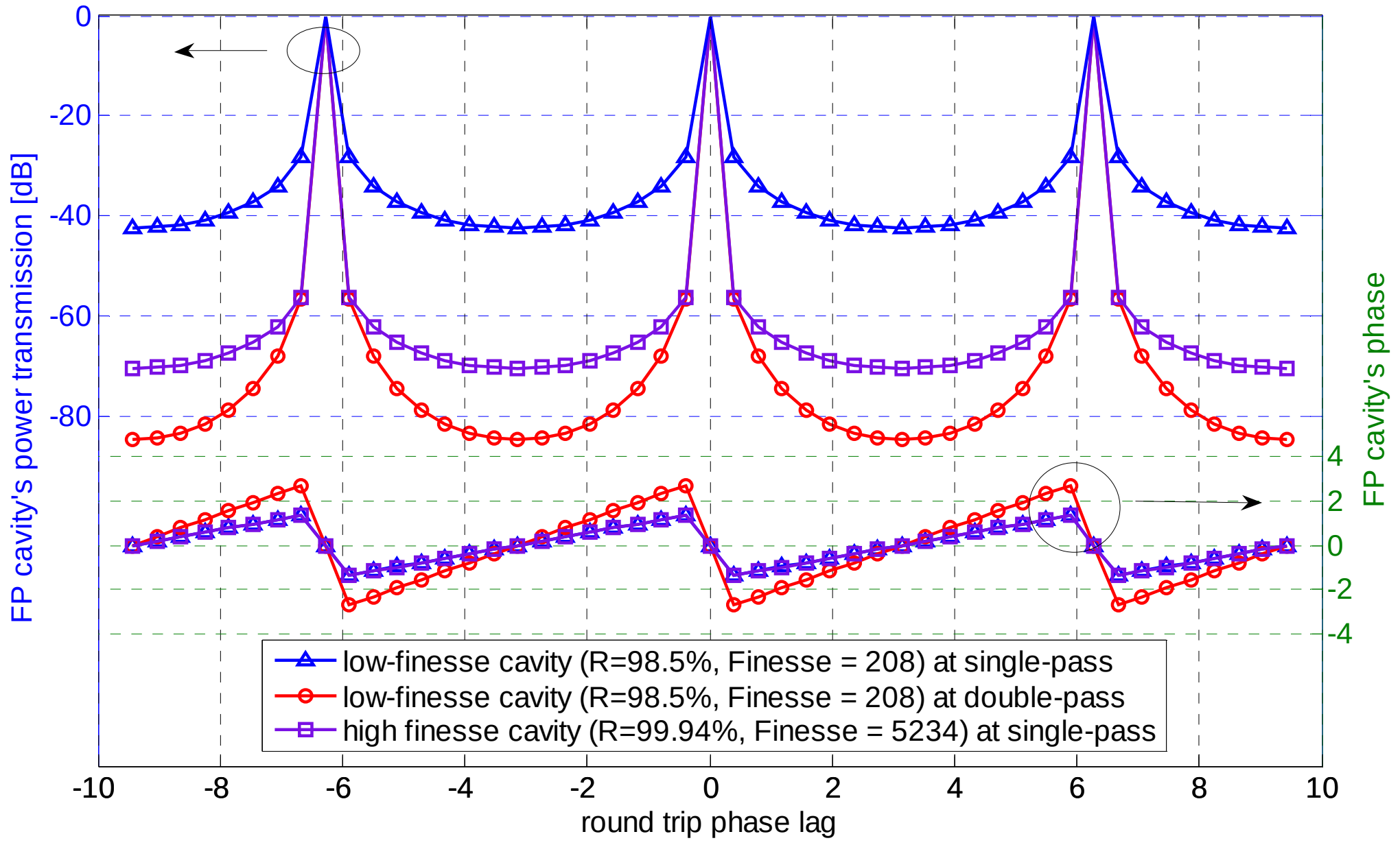

Fig. 2. Comparison of calculated power transmission and phase for three different FP cavities: a low-finesse cavity at single-pass, low-finesse cavity at double-pass, and high-finesse cavity at single-pass. For the simulations, we assume a 1 GHz source comb filtered by a FP cavity with 16 GHz FSR. Note that the cavity phases for the low-finesse cavity and high-finesse cavity, both operating at single-pass, are almost indistinguishable.

The third main component of the proposed broadband astro-comb is a piece of highly nonlinear optical fiber for substantial spectral broadening of the narrowband astro-comb that results from double-pass filtering. To make such an approach practical, power amplifiers might be employed to compensate for filter losses due to the FP cavity and to generate high enough pulse energies at the input of the optical fiber. Chirped-pulse-amplification may also be used to mitigate unfavorable nonlinear effects during power amplification. However, the optical fiber must have a large nonlinearity for the required spectral broadening. Hence, one question naturally arises: how does the nonlinearity affect the side-mode suppression and hence the calibration precision of astronomical spectrographs? In the following sections, we address this question via detailed numerical modeling using realistic parameters — and find a promising result.

### 3. Broadband astro-comb model

In our model of the proposed broadband astro-comb, the source-comb is assumed to emit a train of identical pulses characterized by a repetition-rate $f_{rs}$, which is the comb-line spacing in the spectral domain. The FP filter cavity passes every $M^{th}$ line of the source-comb, thereby increasing the astro-comb line spacing to $f_{ra}$ (i.e., $f_{ra} = Mf_{rs}$, where $M$ is an integer), with $M-1$ suppressed side-modes between adjacent astro-comb lines [Fig. 1(b)]. Such a filtering process can be modeled in the frequency domain using the amplitude transmission of the FP cavity, which in single-pass configuration may be written as

$$t_s = (1-R)/(1-R\times e^{-j\phi}). \qquad (1)$$

Here $R$ denotes the mirror reflectivity, $\phi = 2(\omega/c)n_{air}d + \phi_{mirror}$ accounts for the round-trip phase lag, and $n_{air}$, $d$, and $\phi_{mirror}$ correspond to the air's refractive index, the cavity's physical length and the phase contribution from both cavity mirrors, respectively. The amplitude transmission for a double-pass cavity is simply the square of $t_s$.

In the time domain, the FP cavity filtering process multiplies the source-comb's repetition rate by a factor of $M$; and finite side-mode suppression causes amplitude modulation of the pulse train. See Figure 3. The source comb's repetition-rate defines the amplitude modulation period, $1/f_{rs}$. Since $f_{ra} = Mf_{rs}$, one modulation period covers $M$ pulses.

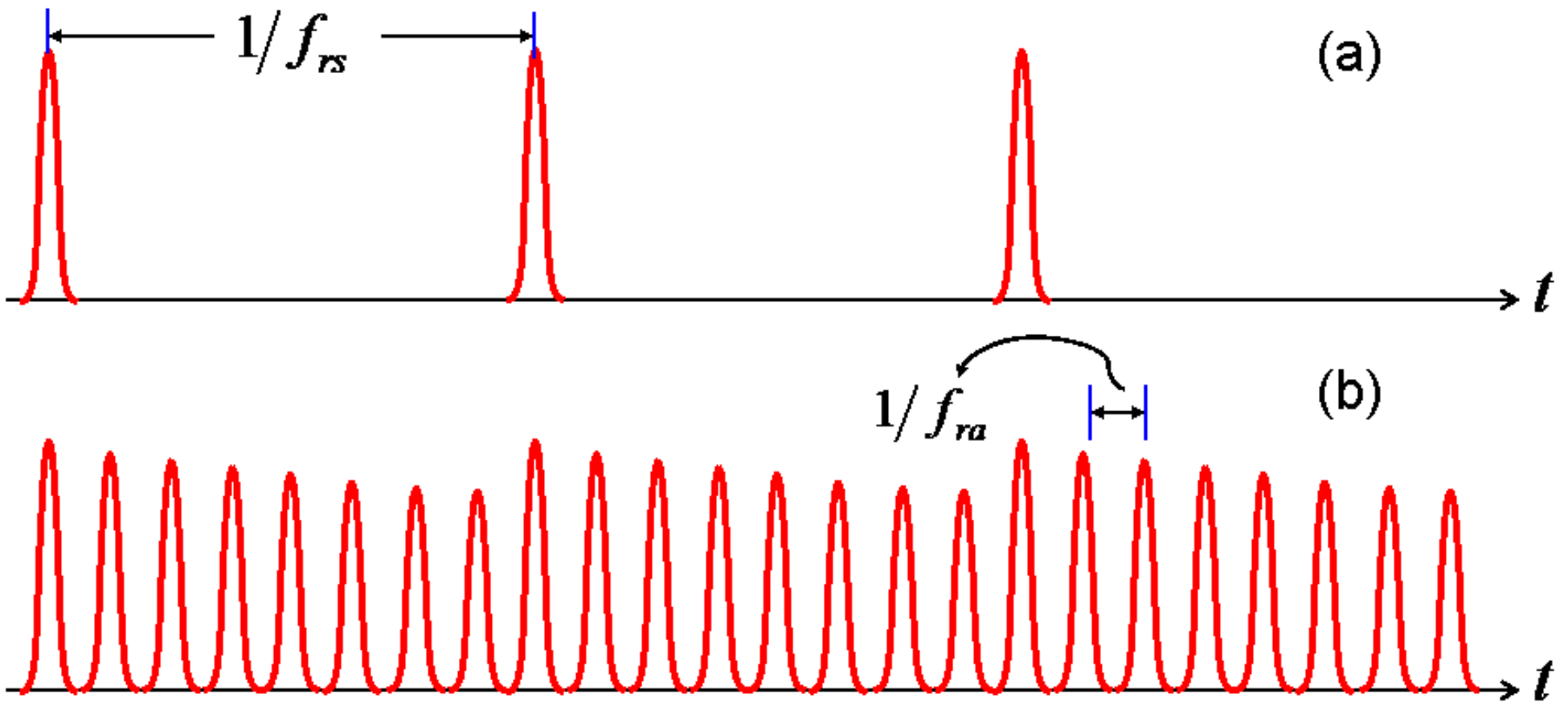

Fig. 3. Schematic to illustrate the time-domain modulation effect of FP cavity filtering of a uniform source-comb pulse train: (a) pulse train before FP cavity filtering; (b) pulse train after FP cavity filtering. In this example, $f_{ra} = 8f_{rs}$, i.e., $M$=8. Note that the amplitudes of these two pulse trains are displayed on different amplitude scales to emphasize the effect of finite FP cavity finesse — and hence finite side-mode suppression and time-domain amplitude modulation.

Nonlinear spectral broadening of ultrashort pulses in an optical fiber can be modeled by the well-known generalized nonlinear Schrödinger (GNLS) equation [7]

$$\frac{\partial A}{\partial z} + \left(\sum_{n=2} \beta_n \frac{i^{n-1}}{n!}\frac{\partial^n}{\partial T^n}\right)A = i\gamma\left(1+\frac{i}{\omega_0}\frac{\partial}{\partial T}\right)\left(A(z,T)\int_{-\infty}^{+\infty} R(t')\left|A(z,T-t')\right|^2 dt'\right) \qquad (2)$$

where $A(z,t)$ denotes the pulse's amplitude envelope. $\beta_n$, $\gamma$, and $\omega_0$ characterize for n-th order fiber dispersion, fiber nonlinearity, and pulse center frequency, respectively. $R(t)$ describes both the instantaneous electronic and delayed molecular responses of fused silica, and is defined as

$$R(t) = (1-f_R)\delta(t) + f_R(\tau_1^2+\tau_2^2)/(\tau_1\tau_2^2)\exp(-t/\tau_2)\sin(t/\tau_1), \qquad (3)$$

where typical values of $f_R$, $\tau_1$, and $\tau_2$ are 0.18, 12.2 fs, and 32 fs, respectively [7].

As a train of ultrashort pulses propagates inside an optical fiber, we can neglect the interaction among pulses if their separation is much larger than the pulse duration. If all the pulses are identical, which holds for the pulse train emitted from an ultrafast laser, one can consider a single pulse as the input to the GNLS equation and simulate its evolution using the split-step Fourier method. To guarantee simulation accuracy, the temporal window $\Delta t$ should be much larger than the pulse duration (but less than the time between pulses); similarly, the simulation's spectral window $\Delta f$ must be larger than the pulse spectral width. The product of $\Delta t$ and $\Delta f$ defines the number of sampling points. Since nonlinear spectral broadening is normally accompanied by temporal stretching, both $\Delta t$ and $\Delta f$ need to be properly set *a priori* if the window size is fixed during the simulation. For example, we might choose $\Delta t$ =1 ps and $\Delta f = 10^{15}$ Hz for an initial 60 fs pulse, which after propagation through a short fiber is stretched to 100-fs with its spectrum broadened to 400-nm ($\approx 1 \times 10^{14}$ Hz for 1.06 μm center wavelength). This selection results in a reasonable sampling number of 1,000, which ensures accurate simulations running at high speed.

Such fast single-pulse simulation, unfortunately, fails in modeling the nonlinear spectral broadening of the proposed astro-comb because the pulse train entering the fiber exhibits slow amplitude modulation caused by finite side-mode suppression of the FP cavity [Figure 3(b)]. Since the modulated pulse train repeats itself every $M$ pulses, one might group $M$ consecutive pulses as a new single pulse and feed it into the GNLS equation. For a 1 GHz source comb, this pulse ensemble is 1 ns long, which implies a sample number as large as 1,000,000 and dramatically slows the simulation. Since adjacent pulses are still separated much further than the pulse duration at the output of the fiber, the absence of interaction among pulses is justified. Therefore, we decompose the long pulse into $M$ individual pulses, with each pulse one centered in a 1-picosecond simulation window. After propagation through the fiber, the spectrally-broadened $M$ pulses are stitched back to recover the amplitude modulated pulse train in one modulation period, from which the astro-comb's performance is evaluated. In comparison to directly simulating the pulse train ensemble, this method reduces simulation time by many orders of magnitude.

## 4. Simulation and results

We simulated the performance of a specific realization of a broadband astro-comb: a source comb constructed from a 1 GHz Yb-fiber laser operating at 1.06 μm; a FP cavity with 16 GHz FSR; and 2 cm SF6-glass photonic crystal fiber (PCF) for spectral broadening. SF6-glass has a nonlinear coefficient that is an order of magnitude larger than fused-silica, which allows significant spectral broadening of low-energy optical pulses using a short segment of fiber. For example, 4 cm of this PCF has been shown to broaden a 20 pJ, 60-fs pulse (the direct output of an Yb-fiber oscillator) into an octave-spanning supercontinuum [8].

As described in the last section, our simulation includes two steps: FP cavity filtering and subsequent spectral broadening. The input to the FP cavity is a train of hyperbolic-secant pulses with 60 fs duration, and a 1 GHz repetition-rate. Fourier transformation of the pulse train multiplied by the FP cavity's amplitude transmission (single-pass or double pass) leads to a narrowband astro-comb spectrum; inverse Fourier transformation of this spectrum yields 16 pulses uniformly distributed in a 1 ns window. In modeling the filtering cavity, we adopt $\phi = 2(\omega / c)d$ in equation (1), i.e., we neglect the GDD from the air and cavity mirrors; this is justified considering that the input pulse has a bandwidth of only 30 nm. The 16 pulses/ns have slightly different pulse energy, and we assume that the strongest one is 60 pJ at the input of the PCF. Therefore, the average power of the pulse train into the PCF is ~ 1 W. Such a power (and energy) level can be easily achieved by linearly amplifying the filtered narrowband comb via an Yb-doped fiber amplifier. Parameters of the SF6-glass fiber for the nonlinear spectral broadening are adapted from Ref. 8, i.e., 1.7 μm for the mode-field diameter and 570 $W^{-1}km^{-1}$ for the nonlinear parameter $\gamma$. The fiber's dispersion is obtained by fitting experimental data (Figure 2 in Ref. 8) with a $6^{th}$-order polynorminal.

To investigate the effect of the FP cavity on side-mode suppression, we compare three filtering schemes: 1) single-pass of source-comb light through a low finesse cavity with $R = 98.5\%$; 2) double-pass through a low finesse cavity with $R = 98.5\%$; and 3) single-pass through a high finesse cavity with $R = 99.94\%$.

Figure 4 summarizes the simulation results for scheme 1. The nonlinearly broadened astro-comb spectrum and its $1^{st}$ side-modes are plotted in Fig. 4(a) at the same scale. The initial narrowband astro-comb spectrum shown in the inset acquires nearly 10 times more bandwith. As expected, this highly-nonlinear process is accompanied by the rapid growth of side-modes. For nonlinear propagation in a 2 cm length of PCF, the dominant effects are self-phase modulation (SPM) and dispersion. SPM, a special type of degenerate four-wave mixing (FWM), causes spectral broadening by redistributing the total power among frequency components. In particular, for initially weak side-modes, nearby strong astro-comb lines act as pumps that provide parametric gain through FWM. It is worth noting that optical parametric amplification is a phase-sensitive nonlinear process [9], in which the phase relations of spectral lines (astro-comb lines or side-modes) determine the power flow among them. Although 1st order side-modes on either side of an astro-comb line experience identical suppression from the FP cavity, these side-modes receive opposite phase shifts relative to the astro-comb line sitting between them, as shown in Fig. 2. It is this phase asymmetry that induces the difference in growth of side-mode power. In other words, the side-mode's phase asymmetry due to FP cavity filtering is translated into amplitude asymmetry via nonlinear spectral processes. Figure 4(a) plots the simulated astro-comb line and upper and lower 1st side-mode power as a function of wavelength, showing that the side-modes account for a considerable portion of the total power. To verify that the large relative difference in side-mode suppression is due to the initial phase, we intentionally switched the phase of the upper and lower 1st side-mode at the PCF's input: the corresponding side-mode power spectra is then broadened by the PCF interchange as expected. Figure 4(b) illustrates the simulated variation of side-mode suppression with wavelength. Within the spectral range of 0.97 -1.17 μm (i.e. full-width at half-maximum of the source-comb), the suppression difference between the upper and lower 1st side-modes is as large as 10 dB [inset of Fig. 4(b)].

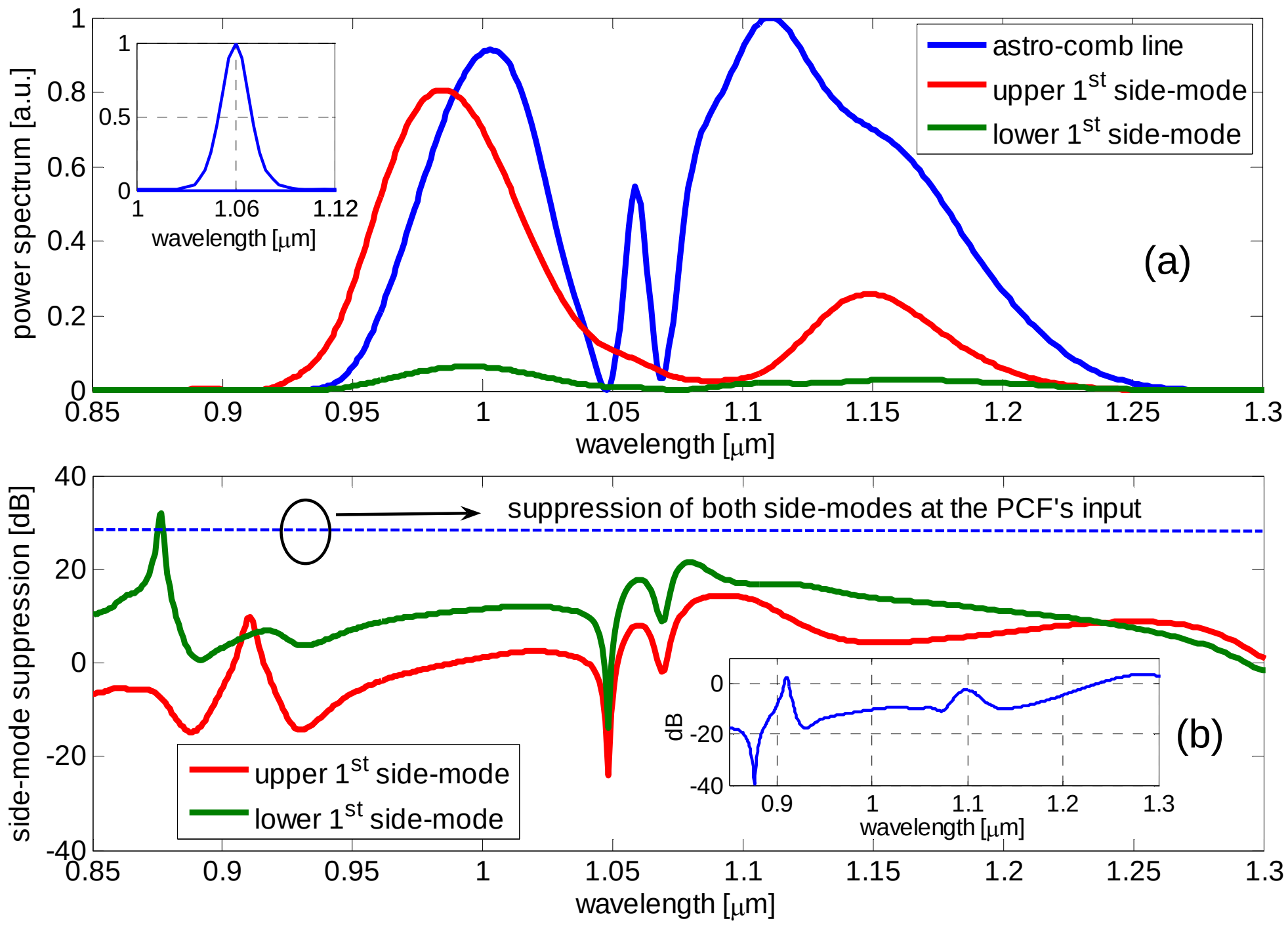

Fig. 4. Simulation results corresponding to single-pass filtering with a low finesse cavity. (a) Broadband astro-comb lines (blue), upper 1st side-mode (red), and lower 1st side-mode (green) at the output of the 2-cm PCF. Inset is the input narrowband spectrum. (b) Suppression of both side-modes after nonlinear spectral broadening. The dashed line indicates suppression at the PCF input. Inset shows the suppression difference between the upper and lower 1st side-modes.

Figure 5 presents simulation results for scheme 2: double-pass of source-comb light through the same low finesse FP cavity. This double-pass technique increases initial side-mode suppression by the FP cavity, relative to the single-pass technique, from 28 dB (34 dB) to 56 dB (68 dB) for the 1st (2nd) side-mode. This greater initial suppression at the PCF input provides much less power in the side-modes for FWM processes, which greatly mitigates phase-asymmetry-to-amplitude-asymmetry translation.

In particular, within a 300 nm 10-dB bandwidth (0.93 -1.23 μm, green curve in Fig. 5(b)), nonlinear processes induce a 10-30 dB degradation of suppression for both the 1st and 2nd side-modes [Fig. 5(a)], leaving their 12-dB relative suppression difference almost unaffected. Importantly, however, the suppression asymmetry is drastically reduced to less than 0.05 dB (0.2 dB) for the upper and lower 1st (2nd) side-modes; this excellent side-mode suppression symmetry is essential for application of astro-combs to precision calibration of astronomical spectrographs.

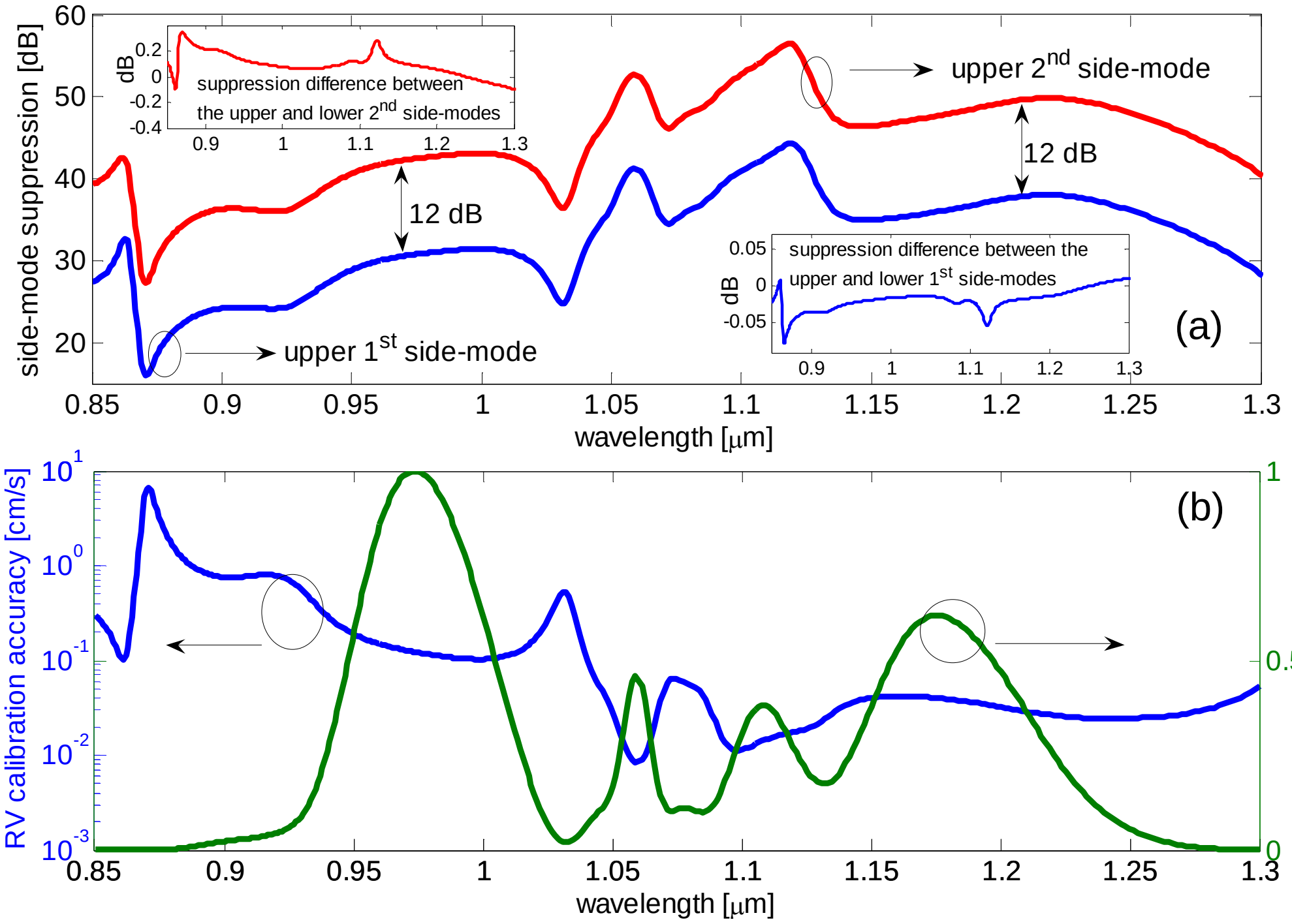

Fig. 5. Simulation results corresponding to double-pass filtering with a low finesse cavity. (a) Suppression of the upper 1st (blue) and 2nd (red) side-modes. Insets show the suppression difference for the 1st (blue) and 2nd (red) side-modes. (b) Broadband astro-comb spectrum (green) and limit on RV calibration accuracy (blue) determined from side-mode suppression and asymmetry.

In particular, spectrograph calibration accuracy depends on both side-mode suppression and symmetry with respect to the nearest main astro-comb lines. These two factors define the center-of-gravity (COG) of astro-comb lines as measured by an astrophysical spectrograph of finite wavelength resolution. A shift between the COG and the true astro-comb line frequency generates a systematic calibration error. The net systematic frequency shift can be calculated as the power-weighted average of the frequency separation between the side-modes and the nearest astro-comb line:

$$\Delta f \approx f_{rs} \sum_{j=1}^{m} j(\rho_j^u - \rho_j^l), \quad (4)$$

where $1/\rho_j^u$ ($1/\rho_j^l$) corresponds to the suppression of the upper (lower) $j^{th}$ side-mode; and $m$ is determined by the spectrograph's resolution. Owing to the stronger suppression of higher-order modes as well as their smaller amplitude asymmetry, it is generally sufficient to set $m = 3$. For studies of astronomical radial velocity (RV) changes using Doppler shift spectroscopy, this systematic calibration error (and associated drifts) limit the spectrograph's RV accuracy (and stability) as follows:

$$\Delta RV = c\frac{\Delta f}{f_0} \approx c\frac{f_{rs}}{f_0} \sum_{j=1}^{m} j(\rho_j^u - \rho_j^l), \quad (5)$$

where $f_0$ and $c$ denote the astro-comb line's frequency and speed of light. As shown in Fig. 5(b), the systematic shift of radial velocity for the proposed broadband astro-comb is $\Delta RV$ <1 cm/s within the 300 nm 10-dB bandwidth; this would bemore than adequate for studies of rocky, Earth-like planets, which require 10-cm/s accuracy on the stellar radial velocity. (Note that there are other requirements to achieve observational sensitivity to stellar $\Delta RV \sim 1$ cm/s, including sufficient photon signal-to-noise from the astronomical source, low guiding error of light through the telescope to the spectrograph, barycentric

corrections, etc.)

Since the narrowband astro-comb has equally spaced lines, FWM occurs for many line combinations during nonlinear spectral broadening. For example, parametric amplifying of the upper 1st side-mode includes participation of: the lower 1st side-mode and the nearest astro-comb line; the nearest astro-comb line and the upper 2nd side-mode; and so on. Thus the first side-modes' suppression and their asymmetry must also depend on the initial phase and amplitude of higher-order modes. To verify this effect, we simulated nonlinear spectral broadening of a narrowband astro-comb using the third filtering scheme: i.e., single-pass through a high finesse cavity ( $R = 99.94\%$ ). See Figure 6. This scheme provides the same suppression (56 dB) for the 1st side-modes as is achieved by the double-pass low-finesse cavity of scheme 2; however, the higher-order side-modes are less suppressed. Figure 2 indicates that this high-finesse cavity suppresses the second side-mode by 62 dB, as opposed to 68 dB in the double-pass cavity. It can be seen from the simulation results presented in Fig. 6 that this reduced suppression of high-order side-modes grossly degrades the astro-comb's performance: the side-mode asymmetry is as large as 1 dB. As shown in Fig. 6(b), such an increased asymmetry worsens the RV accuracy by about 2 orders of magnitude, rendering the resultant broadband astro-comb inadequate as a spectrograph wavelength calibrator used in searches for and studies of Earth-like planets.

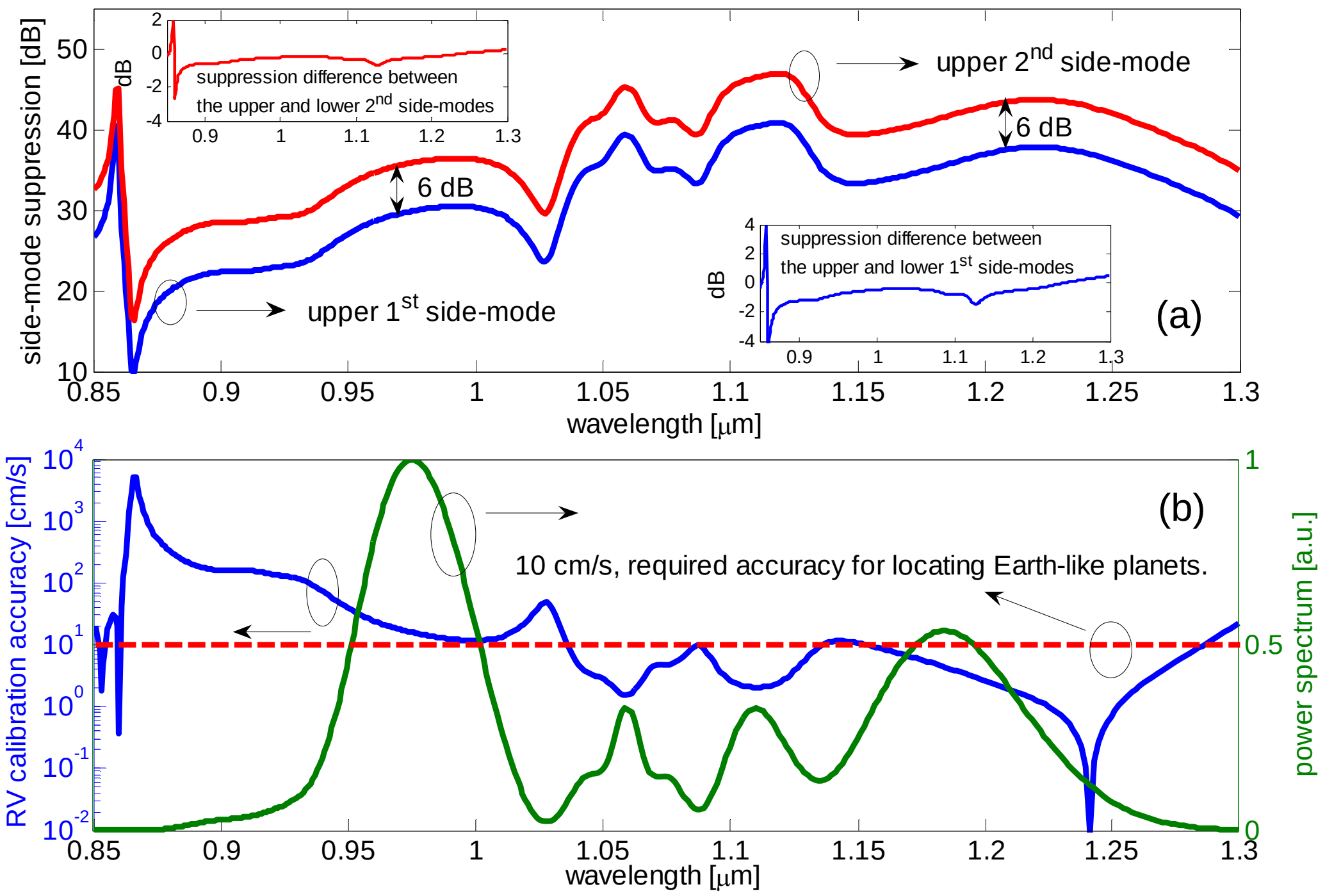

Fig. 6. Simulation results corresponding to single-pass filtering with a high finesse cavity. (a) Suppression of the upper 1st (blue) and the upper 2nd (red) side-modes. Insets show suppression difference for the 1st (blue) and 2nd (red) side-modes. (b) Broadband astro-comb lines (green) and the RV calibration accuracy (blue).

## 5. Discussion and conclusion

In this paper we have studied the interplay of FP cavity filtering schemes and nonlinear spectral broadening by photonic crystal fiber (PCF) in the context of the generation of broadband, high-repetition-rate optical frequency combs suitable for use as wavelength calibrators for astronomical spectrographs ("astro-combs"). Our detailed simulations clearly reveal that the main mechanism for degradation of the filtering cavity's side-mode suppression and amplitude asymmetry is phase-sensitive, multiple four-wave-mixing FWM among astro-comb lines and their side-modes. This degradation results from the high sensitivity of nonlinear spectral broadening to the spectrum of the input signal. Note that related effects have been extensively studied in the context of supercontinuum generation [10-12]. In these related scenarios, the input is usually a train of identical pulses contaminated by low frequency relative intensity noise, which is then amplified and converted to phase noise by the supercontinuum process, leading to reduced coherence of the resulting pulse train. In contrast, in our case the FP filtering cavity is periodically loaded with pulses at the fundamental repetition rate of the source comb, which is a deterministic process (i.e., not noise).

In summary, we have proposed a new approach to generate broadband astro-comb light by using highly nonlinear optical fiber for spectral broadening of a narrowband astro-comb spectrum, which in turn is generated by cavity filtering of an input source-somb spectrum. We demonstrated the feasibility of this approach with detailed numerical simulations fed with realistic parameters. We found that nonlinear spectral broadening can aggravate the side-mode suppression provided by the filtering cavity, and also introduce side-mode amplitude asymmetry; therefore resulting in reduced accuracy for spectrograph calibration. Fortunately, we also found that these deleterious consequences can be mitigated using a double-pass filtering cavity (or equivalently, single-pass through two cascaded cavities), which efficiently suppresses the side-modes prior to spectral broadening. Numerical simulations demonstrated that the proposed astro-comb is able to cover a 300-nm bandwidth and provide spectrograph calibration accurate to 1 cm/s in stellar radial velocity measurements, much less than the 10-cm/s accuracy requirement for the search for and characterization of Earth-like planets.

**Acknowledgements:**

This work was funded under NASA award number NNX09AC92G and NSF grants AST-0905214 and 0905592.